\documentclass[11pt]{article}
\usepackage{graphicx}

\usepackage{epstopdf}
\usepackage{tikz}
\usepackage{amsmath,amssymb,amscd,latexsym,amsthm,mathrsfs}

\ifx\pdfoutput\undefined\else

\renewcommand{\author}[1]{\large\rm #1\\ \bigskip}
\newcommand{\address}[1]{{\normalsize\it #1\\}\bigskip}
\renewcommand{\title}[1]{\bigskip\bigskip\Large\bf #1\bigskip\bigskip\\}

\newcommand{\bexp}{b}

\begin{document}
\vglue .3 cm
\begin{center}

\title{A series test of the scaling limit of self-avoiding walks}
\author{              Anthony J. Guttmann\footnote[1]{email:
                {\tt tonyg@ms.unimelb.edu.au}}}
\address{ ARC Centre of Excellence for\\
Mathematics and Statistics of Complex Systems,\\
Department of Mathematics and Statistics,\\
The University of Melbourne, Victoria 3010, Australia}
\author{              Jesper L. Jacobsen\footnote[2]{email:
                {\tt jesper.jacobsen@ens.fr}}}
\address{ 
LPTENS, \'Ecole Normale Sup\'erieure, 24 rue Lhomond, 75231 Paris, France\\
Universit\'e Pierre et Marie Curie, 4 place Jussieu, 75252 Paris, France}

\end{center}
\setcounter{footnote}{0}
\vspace{5mm}

\begin{abstract}
It is widely believed that the scaling limit of self-avoiding walks (SAWs) at the critical temperature is (i) conformally invariant, and (ii) describable by Schramm-Loewner Evolution (SLE) with parameter $\kappa = 8/3.$ We consider SAWs in a rectangle, which originate at its centre and end when they reach the boundary. We assume that the scaling limit of SAWs is describable by ${\rm SLE}_\kappa,$ with the value of $\kappa$ to be determined. It has previously been shown by Guttmann and Kennedy \cite{GK13} that, in the scaling limit, the ratio of the probability that a SAW hits the side of the rectangle to the probability that it hits the end of the rectangle, depends on $\kappa.$ By considering rectangles of fixed aspect ratio 2,  and also rectangles of  aspect ratio 10, we calculate the probabilities exactly for larger and larger rectangles. By extrapolating this data to infinite rectangle size, we obtain the estimate $\kappa = 2.66664 \pm 0.00007$ for rectangles of aspect ratio 2 and  $\kappa = 2.66675 \pm 0.00015$ for rectangles of aspect ratio 10. We also provide numerical evidence supporting the conjectured distribution of SAWs striking the boundary at various points in the case of rectangles with aspect ratio 2. 
\end{abstract}
 
\section{Introduction}
\begin{figure}[ht!]
\begin{center} 
\includegraphics[width=5cm]{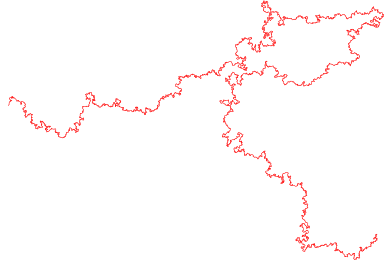}
\caption{A typical two-dimensional SAW of $2^{25}$ steps on the square lattice --
 courtesy of Nathan Clisby.}\label{fig:saw} 
\end{center}
\end{figure}

A self-avoiding walk (SAW)  of length $n$ 
on a periodic graph or lattice                      ${\mathcal L}$ 
is a sequence of distinct
vertices $w_0,w_1,
\ldots ,w_n$ in ${\mathcal L}$ 
such that each vertex is a nearest
neighbour of its predecessor. In Figure~\ref{fig:saw} a very long walk of $2^{25}$ steps is shown (generated by a Monte Carlo algorithm \cite{C10}).

\noindent
Consider now those SAWs starting at the centre of an $L \times W$ rectangle and ending when they hit the boundary. We take the scaling limit of SAW, at the critical point, in the usual way (described, for example, in \cite{GK13}). Let $p_L$ be the probability that a randomly chosen SAW, in the scaling limit, hits the side of the rectangle before the end, and let $p_W=1 - p_L$ be the probability that it hits the end before the side. In Guttmann and Kennedy \cite{GK13} the probability ratio $p_L/p_W$ was calculated, under the assumption that the scaling limit is describable by ${\rm SLE}_{8/3}.$ For a rectangle of aspect ratio $r=L/W,$ with parameter $b,$ where $b=\frac{3}{\kappa}-\frac{1}{2},$ they found, asymptotically,
$$\frac{p_W}{p_L}=\frac{2^{2b+1}\Lambda}{b} e^{-b\pi r/2}\left [ 1 + \frac{\Lambda 2^{b+1}} {b \sin\left ( \frac{\pi b}{2} \right )}e^{-b\pi r/2} + 4(b-1+2\Lambda)e^{-\pi r/2}+ O(e^{-b\pi r})\right ],$$ for $0 < b < 1,$ where $\Lambda = \left (\frac{\Gamma \left ( \frac{1+b}{2} \right ) }{\Gamma \left ( \frac{b}{2} \right ) }\right )^2.$ The exact result is given as the ratio of two integrals (\ref{rat}), and we give an outline of their calculation in the following section.

In this work, we have considered all $2n \times 4n$ rectangles of aspect ratio 2 of size $2 \times 4$ up to $18 \times 36,$ and have generated all SAW starting at the origin and ending on the boundary of the rectangles. We have also considered all $2n \times 20n$ rectangles of aspect ratio 10 of size $2 \times 20$ up to $14 \times 140$. We have also developed a refined enumeration scheme in which the full hitting distribution is
recorded; we give the results here for rectangles of aspect ratio 2. The relevant geometry is shown in Figure~\ref{fig:lattice}. We obtained separately
the generating functions for SAWs starting at the centre of the rectangle and exiting at any prescribed point
on its boundary. This has been compared to that calculated by conformal mapping, assuming that ${\rm SLE}_{8/3}$ is the appropriate scaling limit.

The data was generated using 
a transfer matrix formalism, utilising symmetry and similar techniques to minimise the computational complexity. Details of the methods used are given in Section \ref{series}. In addition to the position of the end-points of the walks on the boundary, the length (number of steps) was also retained.

In \cite{K01} a Monte Carlo study was made by Kennedy, who compared certain random variables associated with SAWs in a half-plane with those predicted assuming that the scaling limit of SAWs is given by ${\rm SLE}_{8/3}.$ Excellent  agreement was found in a visual comparison of two graphical plots. Our work differs in nature in that we predict a specific value of $\kappa$, assuming that 
the scaling limit of SAWs is given by ${\rm SLE}_{\kappa}.$

 \subsection{Conformal mapping}

Let $D$ be a bounded, simply connected domain in the complex plane containing $0.$ We are interested in paths in $D$ starting at $0$ and ending on the boundary of the domain. Initially we will consider random walks, later self-avoiding walks.  We can discretize the space with a lattice of lattice spacing $\delta.$ In both the random walk case and the SAW case we are then interested in the scaling limit $\delta \to 0.$ For random walks the scaling limit is Brownian motion, stopping when it hits the boundary of $D.$ The distribution of the end-point is harmonic measure. If the domain boundary is piecewise smooth, then harmonic measure is absolutely continuous with respect to arc length along the boundary \cite{R16}. Let $h_D(z)$ denote the density with respect to arc length (often called the Poisson kernel). If $f$ is a conformal map on $D$ that fixes the origin, and such that the boundary of $f(D)$ is also piecewise smooth, then the conformal invariance of Brownian motion implies that the density for harmonic measure on the boundary of $f(D)$ is related to the boundary of $D$ by 
\begin{equation}\label{cirw}
h_D(z)=|f'(z)|h_{f(D)}(f(z)).
\end{equation}
In the case of the SAW, Lawler, Schramm, and Werner \cite{LSW04} predicted that the corresponding density of the probability measure $\rho(z)$  transforms under conformal maps as
\begin{equation}\label{cisaw}
\rho_D(z)=c|f'(z)|^{\bexp}\rho_{f(D)}(f(z)),
\end{equation}
where $\bexp=5/8$ and the constant $c$ is required to ensure that $\rho_D(z)$ is a probability density. If one starts the random walk or the SAW at the centre of a disc, then the hitting density on the circle will be uniform. So the above equations determine the hitting density for any simply connected domain, by virtue of the Riemann mapping theorem. Simulations in \cite{DGKLP11}, \cite{KL11} and \cite{K12} provide strong support for the conjectured behaviour. (Eq. (\ref{cisaw}) is correct for domains whose boundary consists only of vertical and horizontal line segments. For general domains there is a lattice effect that persists in the scaling limit that produces a factor that depends on the angle of the tangent to the boundary that must be included \cite{KL11}.) 

 The solution of the problem for random walks by conformal maps is described in \cite{BLWW04}, where a conformal map from the unit disc to an $a \times c$ rectangle is given by the Schwarz-Christoffel formula. Our approach also uses conformal maps, but we instead use a map between the upper half-plane and a rectangle, where the mapping is again given by a Schwarz-Christoffel transformation. For $\alpha > 1,$ let
$$f(z) = \int_0^z \frac{d\xi}{\sqrt{1-\xi^2}\sqrt{\alpha^2 - \xi^2}}.$$ $f(z)$
 is a Schwarz-Christoffel transformation that maps the upper half plane to a rectangle.
The rectangle has one edge along the real axis and $0$ is a midpoint of this side. So the corners
can be written as $\pm a/2$ and $ic \pm a/2$ where $a, c > 0$ are the length of the horizontal and vertical
edges, respectively. We have
$$f(1) = a/2, \,\, f(-1)=-a/2, \,\, f(\alpha) = a/2 + ic, \,\, f(-\alpha) = -a/2+ic, \,\, f(0) =0.$$

So
$$ a=\int_{-1}^1 \frac{dx}{\sqrt{1-x^2}\sqrt{\alpha^2 - x^2}}, \,\,\, c =\int_{1}^\alpha \frac{dx}{\sqrt{x^2-1}\sqrt{\alpha^2 - x^2}}.$$ 
We note that $$ a =\frac{2}{\alpha}{\bf K} \left ( \frac{1}{\alpha} \right ), \,\, c=\frac{1}{\alpha}{\bf K} \left ( \frac{\sqrt{\alpha^2-1}}{\alpha} \right ), \,\, \alpha > 1.$$
Here ${\bf K}(x)$ is the complete elliptic integral of the first kind. By dilation invariance we only need concern ourselves with the aspect ratio $r=a/c.$ So given an aspect ratio
$r$, we have to find $\alpha$ such that

$$ r = \frac{\int_{-1}^1 (1-x^2)^{-1/2}(\alpha^2 - x^2)^{-1/2} dx}{\int_{1}^{\alpha} (x^2-1)^{-1/2}(\alpha^2 - x^2)^{-1/2} dx} = \frac{2 {\bf K} \left ( \frac{1}{\alpha} \right )}{{\bf K} \left ( \frac{\sqrt{\alpha^2-1}}{\alpha} \right )}.$$

From the properties of elliptic integrals, it follows \cite{GK13} that
\begin{equation}\label{alfa}
\sqrt{\alpha} = \frac{\theta_3(e^{-2\pi/r})}{\theta_2(e^{-2\pi/r})},
\end{equation}
where $\theta_j(q)=\theta_j(0,q)$ is the Jacobi theta function.
Evaluating this in one's favourite algebraic package gives the required value of $\alpha$ for any $r \ge 1$ instantly.

Alternatively, we note that for an aspect ratio of 10, $\alpha$ is very close to 1. We can achieve very high  accuracy by expanding the
ratio of the above integrals around $\alpha=1$, and find
$$r=\frac{1}{\pi}\left ( 4\log(2\sqrt{2})-2\log(\alpha-1)+(\alpha-1)-\frac{3}{8}(\alpha-1)^2+\frac{5}{24}(\alpha-1)^3+O(\alpha-1)^4 \right ).$$
Solving this numerically for $r=10$ gives $\alpha=1.00000120561454706472212\ldots.$
To leading order one obtains $$\alpha\approx 1+8 e^{-\pi r/2}+32 e^{-\pi r},$$ which for $r=10$ gives 19 digit accuracy. 
 
 An alternative approach  is to note that $$r=\frac{{2\bf K'}\left (\frac{\sqrt{\alpha^2-1}}{\alpha}\right )}{{\bf K}\left (\frac{\sqrt{\alpha^2-1}}{\alpha}\right )},$$ so that when  $r^2 = 0 \mod 4,$ one has $$\frac{\sqrt{\alpha^2-1}}{\alpha}=k_{r^2/4}.$$ Here $k_n$ are {\em singular moduli,} that is, elliptic moduli for which $\frac{{\bf K}(k_n)}{{\bf K'}(k_n)} = \frac{1}{\sqrt{n}}.$ From a table of singular moduli, for example \cite{sing_mod}, one finds for $r=10$ that
 $$\alpha = \frac{\sqrt{2+24\sqrt{161\sqrt{5}-360}}}{1+12\sqrt{161\sqrt{5}-360}}.$$ For aspect ratio $r=2,$ one finds $\alpha=\sqrt{2}.$

These considerations establish $\alpha,$ a parameter of the conformal map, both directly and asymptotically. To proceed further we need to determine the behaviour of the hitting density of paths from the origin to the boundary of the rectangle.

First, note that the preimage of the center of the rectangle will, by symmetry, be on the imaginary axis, so write it as $di.$ In \cite{GK13} geometrical arguments are given to show that $d=\sqrt{\alpha.}$ An alternative algebraic derivation is the following:
From the Schwarz-Christoffel mapping $f(z)=\int_0^z \frac{d\xi}{\sqrt{1-\xi^2}\sqrt{\alpha^2-\xi^2}},$ set $u=-i \xi,$ giving
$ic/2=f(id)=i\int_0^d \frac{du}{\sqrt{1+u^2}\sqrt{\alpha^2+ u^2}}.$
This integral is an (incomplete) elliptic integral of the second kind, from which follows $d=\sqrt{\alpha}.$

As shown in \cite{GK13},
for both the random walk and the SAW in the half plane starting at $id,$ the (unnormalized)
hitting density along the real axis is $(x^2 + d^2)^{-\bexp} = (x^2 + \alpha)^{-\bexp}$.  Hence the hitting density $\rho_R$ for walks in a rectangle starting at the center is
\begin{equation}\label{rh}
(x^2 + \alpha)^{-\bexp} \propto |f'(x)|^\bexp \rho_R(f(z)).
\end{equation}

The ratio of probabilities of a walk first hitting an end of the rectangle to that of a walk first hitting a side is just  the ratio of the integral of the hitting density $\rho(z)$ along a vertical edge to the integral along a horizontal edge,
$$\frac{\int_{0}^{c} \rho_R(a/2+iy) dy}{\int_{-a/2}^{a/2} \rho_R(x) dx}.$$
By a change of variable, setting $u = f^{-1}(x)$ in the denominator and $u=f^{-1}(a/2+iy)$ in the numerator, and
recalling that $f'(u) = (1-u^2)^{-1/2}(\alpha^2 - u^2)^{-1/2},$  the ratio of probabilities of a first hit on the vertical side to a first hit on the horizontal side, $R(\alpha,\bexp)$, is
\begin{equation}\label{rat}
R(\alpha,\bexp) = \frac {\int_{1}^{\alpha} (u^2+\alpha)^{-\bexp}(u^2-1)^{(\bexp-1)/2}(\alpha^2 - u^2)^{(\bexp-1)/2} du}{\int_{-1}^{1} (u^2+\alpha)^{-\bexp}(1-u^2)^{(\bexp-1)/2}(\alpha^2 - u^2)^{(\bexp-1)/2} du}.
\end{equation}
These integrals cannot be solved in closed form. Both Mathematica and Maple can be used to evaluate them numerically to any reasonable accuracy. However for our purposes it would be useful to have a reasonably accurate asymptotic representation.

We will calculate the leading order and first correction term in the quotient (\ref{rat})
 where $b \in (0,1]$ and  $\alpha=1+c,$ with $|c| \ll 1.$ Here $\alpha$ is related to the aspect ratio $r$ of a rectangle
by (\ref{alfa}). 
Consider the numerator first. One has
$$N(\alpha,b)=\int_1^\alpha (u^2+\alpha)^{-b}(\alpha^2-u^2)^{(b-1)/2}(u^2-1)^{(b-1)/2} du.$$
Now, since $u=1+ \epsilon$ with $\epsilon \ll 1,$
\begin{eqnarray}
(u^2+\alpha) \approx & (2u+c), \nonumber \\
 (\alpha^2-u^2)=&(\alpha-u)(\alpha+u) \approx (2+c+(u-1))(\alpha-u), \\ \nonumber
(u^2-1)=&(u+1)(u-1) = (2+(u-1))(u-1).
\end{eqnarray}
 So
$$N(\alpha,b) \approx 2^{b-1}\cdot \left (1+\frac{c(b-1)}{4} \right )\cdot \int_1^\alpha (2u+c)^{-b}(\alpha-u)^{\frac{b-1}{2}}(u-1)^{\frac{b-1}{2}} du$$
$$+ \, 2^{b-2}\cdot  \int_1^\alpha (2u+c)^{-b}(\alpha-u)^{\frac{b-1}{2}}(u-1)^{\frac{b+1}{2}} du.$$
Set $u=1+tc.$ Then $du=c\cdot dt,$ and
\begin{eqnarray}
N(\alpha,b) \approx & \frac{c^b}{2}\left (1+\frac{c(b-1)}{4} \right )\left ( \int_0^1\left (1+\frac{b-1}{2}tc\right ) [t(1-t)]^{\frac{b-1}{2}} dt - \frac{bc}{2} \cdot \int_0^1 (1+2t)[t(1-t)]^{\frac{b-1}{2}} dt\right ) \nonumber \\
= &\left (1+\frac{c(b-1)}{4} \right )\left ( (\alpha-1)^b \frac{\sqrt{\pi} \cdot \Gamma \left ( \frac{1}{2}+\frac{b}{2} \right ) }{2^{b+1} \cdot \Gamma \left ( 1+\frac{b}{2} \right ) }(1+\frac{b-1}{4}c)- (\alpha-1)^{b+1} \frac{b \cdot \sqrt{\pi} \cdot \Gamma \left ( \frac{1}{2}+\frac{b}{2} \right ) }{2^{b+1} \cdot \Gamma \left ( 1+\frac{b}{2} \right ) } \right ) \nonumber \\
=&(\alpha-1)^b \frac{\sqrt{\pi} \cdot \Gamma \left ( \frac{1}{2}+\frac{b}{2} \right ) }{2^{b+1} \cdot \Gamma \left ( 1+\frac{b}{2} \right ) }\left [1 - \frac{(b+1)}{2}(\alpha-1)\right ]. 
\end{eqnarray}
Next, consider the denominator:
$$F(\alpha,b)=2\int_0^1 (\alpha+u^2)^{-b}[(1-u^2)(\alpha^2-u^2)]^{(b-1)/2} du $$
Again, let $\alpha=1+c,$ where $|c| \ll 1.$ One has
$$(1+c+x^2)^{-b} \approx (1+x^2)^{-b} \left ( 1 - \frac{bc}{1+x^2} \right ),$$
and $$(\alpha^2-x^2) \approx (1-x^2+2c), $$ both to first order in $c.$
So with $t=2c,$ $F(\alpha,b) \approx 2(I(b,t) -bcI(b+1,t)),$ where
$$I(b,t) = \int_0^1 (1+x^2)^{-b}[(1-x^2)(1-x^2+t)]^{(b-1)/2} dx.$$
(Note that the first argument refers to the power of $(1+x^2)$ only, not the other occurrences of $b$).
Next, we take the Mellin transform
$$\int_0^\infty \frac{t^{s-1}}{(1-x^2+t)^{(1-b)/2}} dt = B \left (s,\frac{1-b}{2}-s \right )(1-x^2)^{s+(b-1)/2}.$$
So the Mellin transform of $I(b,t)$ is
$$B \left (s,\frac{1-b}{2}-s \right ) \int_0^1 (1+x^2)^{-b}(1-x^2)^{b+s-1} dx =$$
$$\frac{\sqrt{\pi}}{2} B \left (s,\frac{1-b}{2}-s \right )
\frac{\Gamma(b+s)}{\Gamma(\frac{1}{2}+b+s)} {_2F_1} \left( \frac{1}{2},b;\frac{1}{2}+b+s;-1\right ).$$
Hence the inverse transform is
$$I(b,t) = \frac{\sqrt{\pi}}{2\Gamma \left ( \frac{1-b}{2}\right )} \int_{d-i\infty}^{d+i\infty} \frac{ds}{2\pi i} t^{-s} \frac{\Gamma(s)\Gamma(b+s)\Gamma\left ( \frac{1-b}{2}-s \right )}{\Gamma\left ( \frac{1}{2}+b+s \right )}{_2F_1} \left( \frac{1}{2},b;\frac{1}{2}+b+s;-1\right ),$$
where $0 < d < (1-b)/2.$ Closing the contour to the left, as we want a small $t$ expansion, we see there are poles coming from the Gamma functions. The first occurs at $s=0,$ and the second at $s=-b.$ Taking the residue at the poles we get
$$I(b,t) \approx \frac{\sqrt{\pi}\Gamma(b)}{2\Gamma \left ( b+\frac{1}{2}\right )} {_2F_1} \left( \frac{1}{2},b;\frac{1}{2}+b;-1\right )-2^{-b}c^b \frac{\Gamma(-b)\Gamma\left (\frac{1+b}{2} \right )}{\Gamma\left (\frac{1-b}{2} \right )}.$$
Another term comes from $I(b+1,t),$ and this gives a term
$$-b(\alpha-1)\frac{\sqrt{\pi}\Gamma(b)}{2\Gamma \left ( b+\frac{1}{2}\right )} {_2F_1} \left( \frac{1}{2},1+b;\frac{1}{2}+b;-1\right ).$$
We can eliminate the hypergeometric functions, as they can be expressed in terms of Gamma functions, and then simplified.  In this way we obtain
$$F(\alpha,b) \approx \frac{\sqrt{\pi}\Gamma \left (\frac{b}{2}\right )}{\Gamma \left ( \frac{b+1}{2}\right )}+ (\alpha-1)^b \cdot \frac{\Gamma(-b)\Gamma \left (\frac{1+b}{2}\right )}{\Gamma \left ( \frac{1-b}{2}\right )}-(\alpha-1)\cdot \frac{b\sqrt{\pi}}{4} \left [ \frac { \Gamma \left ( \frac{b}{2} \right )}{\Gamma \left ( \frac{b+1}{2} \right )}  + \frac { \Gamma \left ( \frac{b+1}{2} \right )}{\Gamma \left ( \frac{b}{2}+1 \right )} \right ].$$

We can write the ratio $$\frac{N(\alpha,b)}{F(\alpha,b)} = A \cdot c^b \cdot [1 + Bc +Dc^b + O(c^{2b})] =$$
$$=A8^be^{-b\pi r/2} \left ( 1 + D8^b e^{-b\pi r/2} + (4b+8B)e^{-\pi r/2} + O(e^{-b \pi r}) \right ).$$
Here we have used the asymptotic expansion (\ref{alfa}) of $\alpha-1$ as an expansion in $\alpha,$ and the parameters introduced above are
 $$\Lambda = \frac{\Gamma \left ( \frac{1+b}{2} \right ) ^2}{\Gamma \left ( \frac{b}{2} \right ) ^2},$$ and
$$ A =\frac{\Lambda}{b \cdot 2^{b}},$$
$$B = -\frac{3b}{4}-\frac{1}{2}-\frac{\Lambda}{2},$$
$$D = \frac{\Lambda}{b\cdot 2^b \cdot \sin(\pi b/2)}.$$

Then 
\begin{equation}\label{asympt}
{\tilde R}(r,b)=\frac{2^{2b+1}\Lambda}{b\, e^{b\pi r/2}} \left [ 1 + \frac{\Lambda 2^{2b+1} e^{-b\pi r/2}}{b \sin\left ( \frac{\pi b}{2} \right )} + 4(b-1+2\Lambda)e^{-\pi r/2}+ O(e^{-b\pi r})\right ],
\end{equation}
 for $0 < b < 1.$

For $r=10$ and $b=5/8$ the leading term gives 4 significant digits accuracy (where we are comparing to the value obtained by numerical integration with 50 digit accuracy), while including the next terms gives 8 significant digits. This is more than sufficient for our purpose here. However for $r=2$ (with the same value of $b$) the expansion is much more slowly convergent, and the above expression provides the result to an accuracy of merely 1 part in 20. While we could calculate further terms in the asymptotic expansion, we prefer to simply evaluate the integral expression (\ref{rat}) numerically.

We now use these results to calculate the value of $b$ characterising the scaling limit of SAWs.  We do this by generating all SAWs in a rectangle of given aspect ratio $r$, centred at the origin, and of size $2n \times 2rn,$ for $n = 1,2,\ldots n_{\rm max},$ where $n_{\rm max}$ refers to the largest possible rectangle we can use given our computer resources. 

In the next section we describe the efficient generation of these SAWs, using 
 transfer matrices and efficient use of symmetry and the elimination of unnecessary starting points. In the following section we analyse the data thus produced, by calculating and extrapolating the ratio $R_n(r)$ for two representative values of $r,$ which were $r=2$ and $r=10.$ From these estimates of $R_\infty(r)$ we use the asymptotic expansion above to estimate $b,$ in the case of aspect ration 10, while for $r=2$ the comparison is made with numerical integration values from (\ref{rat}).

Furthermore, we can compare the hitting density distribution function, as calculated from our largest rectangle, to that predicted in the scaling limit. With $b = 5/8,$ and for rectangles of aspect ratio 2, it will be seen that the two distribution functions are almost indistinguishable.

\section{Enumeration of SAWs hitting the short and long sides of the rectangle}\label{series}

We first describe the computation of the generating functions of a SAW starting at the
centre of the strip and terminating, respectively, the first time it hits either one of the 
rectangle's two short sides, or one of its two long sides. The rectangle is supposed
oriented so that the short sides are horizontal and the long sides vertical.

As usual in enumeration studies the basic ingredient is a transfer matrix acting
on an appropriate set of states. Each state comes with a weight, which is a polynomial
with integer coefficients in a variable $x$ that is interpreted as the fugacity for one
step of the SAW. The transfer matrix is maximally decomposed as a product of
elementary matrices that each adds a small piece (a vertex and its adjacent half-edges) to the
lattice being constructed. The states contain information about how the uppermost row
of half-edges are connected among themselves, or to the boundary of the rectangle, or
to the centre point, through pieces of the partially built SAW. This information---and the
action of the elementary matrices on the states---must be carefully devised so that:
\begin{enumerate}
 \item a single SAW going from the centre to the boundary of the rectangle is constructed; 
 \item the two desired generating functions, describing respectively the hitting of the short and long sides of
           the rectangle, can be disentangled;
 \item no superfluous information is stored.
\end{enumerate}
Points 1 and 2 are essential for the correctness of the algorithm, whereas 3 is only a matter
of efficiency. It cannot be excluded that the precise procedure described below can be shown in
the future not to be optimal, in the sense that a variant algorithm can be constructed that uses
fewer states to achieve the goals 1 and 2.

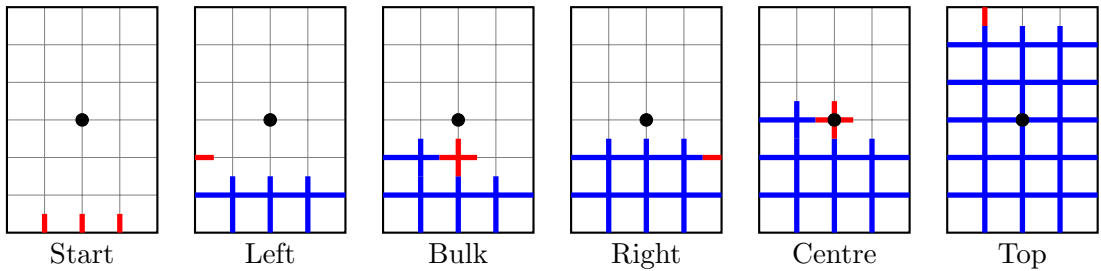
\begin{figure}
\begin{center}

\begin{tikzpicture}[scale=0.5,>=stealth]

 \draw[help lines] (0,0) grid (4,6);
 \draw[black,thick] (0,0) -- (4,0) -- (4,6) -- (0,6) -- cycle;
 \draw[fill] (2,3) circle (1ex);
 \foreach \xpos in {1,2,3}
  \draw[red,line width=2pt] (\xpos,0) -- (\xpos,0.5);
 \draw (2,0) node[below]{Start};

\begin{scope}[xshift=5.0cm]
 \draw[help lines] (0,0) grid (4,6);
 \draw[black,thick] (0,0) -- (4,0) -- (4,6) -- (0,6) -- cycle;
 \draw[fill] (2,3) circle (1ex);
 \foreach \xpos in {1,2,3}
  \draw[blue,line width=2pt] (\xpos,0) -- (\xpos,1.5);
 \draw[blue,line width=2pt] (0,1) -- (4,1);
 \draw[red,line width=2pt] (0,2) -- (0.5,2); 
 \draw (2,0) node[below]{Left};
\end{scope}

\begin{scope}[xshift=10.0cm]
 \draw[help lines] (0,0) grid (4,6);
 \draw[black,thick] (0,0) -- (4,0) -- (4,6) -- (0,6) -- cycle;
 \draw[fill] (2,3) circle (1ex);
 \foreach \xpos in {1,2,3}
  \draw[blue,line width=2pt] (\xpos,0) -- (\xpos,1.5);
 \draw[blue,line width=2pt] (0,1) -- (4,1);
 \draw[blue,line width=2pt] (0,2) -- (1.5,2); 
 \draw[blue,line width=2pt] (1,1.5) -- (1,2.5); 
 \draw[red,line width=2pt] (2,1.5) -- (2,2.5); 
 \draw[red,line width=2pt] (1.5,2) -- (2.5,2); 
 \draw (2,0) node[below]{Bulk};
\end{scope}

\begin{scope}[xshift=15.0cm]
 \draw[help lines] (0,0) grid (4,6);
 \draw[black,thick] (0,0) -- (4,0) -- (4,6) -- (0,6) -- cycle;
 \draw[fill] (2,3) circle (1ex);
 \foreach \xpos in {1,2,3}
  \draw[blue,line width=2pt] (\xpos,0) -- (\xpos,2.5);
 \draw[blue,line width=2pt] (0,1) -- (4,1);
 \draw[blue,line width=2pt] (0,2) -- (3.5,2);
 \draw[red,line width=2pt] (3.5,2) -- (4,2); 
 \draw (2,0) node[below]{Right};
\end{scope}

\begin{scope}[xshift=20.0cm]
 \draw[help lines] (0,0) grid (4,6);
 \draw[black,thick] (0,0) -- (4,0) -- (4,6) -- (0,6) -- cycle;
 \draw[fill] (2,3) circle (1ex);
 \foreach \xpos in {1,2,3}
  \draw[blue,line width=2pt] (\xpos,0) -- (\xpos,2.5);
 \foreach \ypos in {1,2}
  \draw[blue,line width=2pt] (0,\ypos) -- (4,\ypos);
 \draw[blue,line width=2pt] (0,3) -- (1.5,3);
 \draw[blue,line width=2pt] (1,2.5) -- (1,3.5);
 \draw[red,line width=2pt] (2,2.5) -- (2,3.5); 
 \draw[red,line width=2pt] (1.5,3) -- (2.5,3); 
 \draw[fill] (2,3) circle (1ex);
 \draw (2,0) node[below]{Centre};
\end{scope}

\begin{scope}[xshift=25.0cm]
 \draw[help lines] (0,0) grid (4,6);
 \draw[black,thick] (0,0) -- (4,0) -- (4,6) -- (0,6) -- cycle;
 \draw[fill] (2,3) circle (1ex);
 \foreach \xpos in {1,2,3}
  \draw[blue,line width=2pt] (\xpos,0) -- (\xpos,5.5);
 \foreach \ypos in {1,2,3,4,5}
  \draw[blue,line width=2pt] (0,\ypos) -- (4,\ypos);
 \draw[red,line width=2pt] (1,5.5) -- (1,6); 
 \draw[fill] (2,3) circle (1ex);
 \draw (2,0) node[below]{Top};
\end{scope}
 
\end{tikzpicture}
 \caption{Various stages in the transfer matrix construction, here shown for a rectangle of size $4 \times 6$.
 Each stage is labelled by a convenient name and will be explained fully in the main text. The parts of the lattice
 being constructed in any given stage is shown in red, while those already constructed in earlier stages are shown
 in blue. The black dot marks the centre of the rectangle.}
  \label{fig:TM}
\end{center}
\end{figure}

We consider an $L \times W$ rectangle, with $L$ and $W$ even (so that the centre of the rectangle
coincides with a vertex), and $L \le W$ (the transfer matrix then propagates upwards). The states
contain information about the connectivities of the $n$ uppermost half-edges, labelled
$0,1,2,\ldots,n-1$ from left to right, of the partially  constructed lattice. When a row of the lattice is completed,
$n=L-1$ and all half-edges are vertical. At stages where a row is under construction, $n=L$ and one of
the half-edges is horizontal. The lattice is constructed---through the action of the elementary matrices---row
by row, completing the rows from left to right. This is shown in Figure~\ref{fig:TM}.
Each half-edge can be connected to another half-edge, or be in one of the following configurations:
\begin{enumerate}
 \item {\tt EMPTY}, meaning that it is not covered by the SAW;
 \item {\tt SIDE}, when it is connected through the SAW to the left or right boundary of the rectangle;
 \item {\tt BOTTOM}, when its is connected to the bottom (or top) boundary of the rectangle;
 \item {\tt CENTRE}, when it is connected to the centre point.
\end{enumerate}
In addition the system as a whole can be in the following configurations when the SAW has been completed
before the lattice has been completely built:
\begin{enumerate}
\addtocounter{enumi}{4}
 \item {\tt EXIT\_LR} when the SAW went from the centre point to the left or right boundary;
 \item {\tt EXIT\_BT} when it went to the bottom or top boundary.
\end{enumerate}
Being in one of those exit configurations implies that all the half-edges must be empty and remain so until
completion of the lattice. To describe such states it is therefore sufficient to use just $n=1$ label
containing the exit status ({\tt EXIT\_LR} or {\tt EXIT\_BT}).

Imposing the correct constraints throughout the construction of the lattice (see below)
will imply that once the lattice has been completed only the two exit states carry a non-zero
weight. These two weights are precisely the desired generating functions.

We now describe the action on the states of the different elementary matrices that build up the system
(see Figure~\ref{fig:TM}).

In the initial stage, called {\tt Start} in Figure~\ref{fig:TM}, the system is in a superposition of the completely empty state
(with weight $1$), and each of the $n$ states where one of the $n=L-1$ half-edges carries the ${\tt BOTTOM}$
label (with weight $x$) and the others are {\tt EMPTY}. Note that a valid state cannot have more than one half-edge labelled either
{\tt BOTTOM} or {\tt SIDE}, since we aim at constructing a SAW that touches the boundary in only
one single point.

After a row of the lattice has been completed the following operation, called {\tt Left} in Figure~\ref{fig:TM},
adds a horizontal half-edge touching the left boundary of the rectangle. If the system is already in an exit
state it remains so and with the same weight. Otherwise the labels are shifted one unit to the right to make place
for that of the horizontal half-edge. The latter can either be {\tt EMPTY} or, if no other {\tt BOTTOM} or {\tt SIDE}
label exists, be set to {\tt SIDE} in which case the weight is multiplied by $x$ to take account for the extra 
monomer.

The subsequent operations within the row are of the type {\tt Bulk}. The elementary matrix now acts on two labels,
describing the half-edges coming from West and South in the incoming state, and which after the time evolution describe
the half-edges going towards North and East in the outgoing state. An exit state is treated as before. In all
other cases we must ensure that the bulk vertex is adjacent to either two or four occupied edges. Thus, if both
incoming labels are {\tt EMPTY}, the outgoing labels can be {\tt EMPTY} as well, or mutually connected with
an extra weight $x^2$. If only one incoming label is {\tt EMPTY}, the other can go straight through the vertex
or make a turn, in both cases with weight $x$. Finally, if neither of the incoming labels is {\tt EMPTY} there is
a rather large number of possibilities:
\begin{enumerate}
 \item If both labels describe connectivities to other half-edges (i.e., neither is equal to one of the special labels
 {\tt EMPTY}, {\tt SIDE}, {\tt BOTTOM} or {\tt CENTRE} described above), their respective partner half-edges must be
 mutually connected. However, if the connection is among the two incoming half-edges themselves, the
 state must be discarded, since the SAW is not allowed to form a closed loop.
 \item If one of the labels is special, the partner of the non-special label becomes equal to the special label.
 \item Finally, if both incoming labels are special, we have the opportunity of creating an exit state. But
 first a number of precautions must be taken:
 \begin{enumerate}
   \item All the remaining labels in the state need to be {\tt EMPTY}, otherwise the state must be discarded.
   \item One of the incoming labels must be {\tt CENTRE} and the other either {\tt SIDE} or {\tt BOTTOM}
   (otherwise discard the state).
   The latter label then determines whether the exit state is {\tt EXIT\_LR} or {\tt EXIT\_BT}.
  \end{enumerate}
\end{enumerate}
In all those cases the two outgoing labels will be {\tt EMPTY} and the weight coincides with that of the incoming state.

To complete a row we use the operation {\tt Right}. The elementary matrix acts on the rightmost label,
describing the half-edge that will touch the right boundary of the rectangle. The action on exit states is as usual.
If the incoming label is {\tt EMPTY} the state is kept as it is. Otherwise the rightmost half-edge will be occupied,
and we must first make sure that there is not already a label which is {\tt SIDE} or {\tt BOTTOM} (otherwise
discard the state). Next, if the incoming label is {\tt CENTRE} we obtain an {\tt EXIT\_LR} exit state provided
that the remainder of the half-edges are empty. Otherwise, if the incoming half-edge is connected to another
half-edge, the latter will be marked by a {\tt SIDE} label to signify the hitting of the right boundary of the rectangle.
After these considerations the rightmost label is deleted, since we have now completed a row of the lattice
and only $L-1$ labels are required.

Special care must be taken when attaining the centre point. This is achieved by applying the operation {\tt Centre}.
As in the {\tt Bulk} case the action is on two labels, but while in the former case the vertex had degree two or four,
the number of occupied adjacent edges must now be one, since the SAW emanates from the centre point.
Thus, if both incoming labels are {\tt EMPTY} we produce two outgoing states with one of the outgoing labels
being {\tt CENTRE} and the other {\tt EMPTY} (with a weight $x$). States with two occupied ingoing labels
must be discarded. And finally, for states with one ingoing label being {\tt EMPTY} and the other occupied
there are a number of possibilities:
\begin{enumerate}
 \item If the occupied incoming half-edge is connected to another half-edge, the latter acquires the {\tt CENTRE} label.
 \item Otherwise the occupied incoming half-edge is of the type {\tt SIDE} or {\tt BOTTOM}, and we end up in
 an exit state (provided the remainder of the system is empty).
\end{enumerate}

It remains to describe how to finish the rectangle after building the last row. Each half-edge must now be
subjected to the operation {\tt Top} in order to implement the top boundary of the rectangle. The corresponding
elementary matrix acts on a single label (and is in fact very similar to the operation {\tt Right} described above). If that label is {\tt EMPTY} the state is unchanged. If it is {\tt SIDE}
or {\tt BOTTOM} the state is discarded. If it is connected to another half-edge, the latter acquires the
{\tt BOTTOM} label (which is ill-named in this case, since it is the top boundary of the rectangle which is being
touched). And finally, if the incoming label is {\tt CENTRE} we end up in the {\tt EXIT\_BT} state (provided the
remainder of the system is empty).

From a practical point of view, all the labels used to describe a state can be coded on short integers
(say, of the {\tt char} type in {\tt C}) and the coding of the entire state of $n$ half-edges is then an array
of characters. The handling of incoming and outgoing states is then conveniently and efficiently done
by employing standard hashing techniques. The integer coefficients entering the weights will obviously
become very large for even a moderately-sized system, but this is easily dealt with by using modular
arithmetic, i.e., repeating the entire computation modulo various primes and reconstituting the results
from the Chinese remainder theorem.

Our results for the generating functions with various aspect ratios are so lengthy that it makes little sense to
have all them all appear in print. However, the complete results are available in electronic form as
supplementary material to this paper.%
\footnote{This text file provided (named {\tt GJ13.m}) can be processed by {\sc Mathematica} or---maybe after minor changes of
formatting---by any symbolic computer algebra program of the reader's liking.}
For the convenience of the readers who wish to check explicitly the example of the $4 \times 6$ rectangle
shown in Figure~\ref{fig:TM} we give the results for that case.
For the hitting of the long sides of the rectangle we find
\begin{eqnarray}
 2 x^2 + 8 x^3 + 16 x^4 + 12 x^5 + 32 x^6 + 28 x^7 + 52 x^8 +  \nonumber \\ 
 40 x^9 + 76 x^{10} + 56 x^{11} + 
 116 x^{12} + 60 x^{13} + 68 x^{14} \,,
 \label{eq:LR}
\end{eqnarray}
while for the hitting of the short sides we have
\begin{eqnarray}
 2 x^3 + 12 x^4 + 12 x^5 + 24 x^6 + 12 x^7 + 32 x^8 + \nonumber \\
 20 x^9 + 60 x^{10} + 44 x^{11} + 100 x^{12} + 
 28 x^{13} + 56 x^{14} \,.
 \label{eq:BT}
\end{eqnarray}

Note that all the coefficients in this example are even. This is generally true and due to the symmetries
of the rectangle. Explicitly, the mirror symmetry with respect to a horizontal (resp.\ vertical) line going through
the centre point induces a bijection between the SAWs that hit the bottom (resp.\ left) boundary and those that
hit the top (resp.\ right) boundary.

We can take advantage of those symmetries to make a more efficient version of the algorithm. Namely, if
we impose the restriction that all the half-edges incident on the bottom and left boundaries be empty, only the SAWs
hitting the right and top boundaries will be counted. This means that the required generating functions are
simply divided by a factor of two. The reason that we have imposed the constraint on the left boundary
rather than the right, and on the bottom boundary rather than the top, is that the left and bottom boundaries
are those encountered earliest in the time evolution of Figure~\ref{fig:TM}. Imposing the constraints early
on will most efficiently curb the proliferation of unnecessary states.

\begin{table}
\begin{center}
\begin{tabular}{r|rrrr}
 $L$ & $N_1$ & $N'_1$ & $N_2$ & $N'_2$ \\ \hline
     4 & 28 & 18 & 38 & 25 \\
     6 & 162 & 102 & 332 & 211 \\
     8 & 1\,038 & 646 & 2\,844 & 1\,779 \\
   10 & 7\,082 & 4\,376 & 24\,248 & 15\,031 \\
   12 & 50\,448 & 31\,022 & 206\,978 & 127\,557 \\
   14 & 370\,866 & 227\,268 & 1\,772\,136 & 1\,087\,711 \\
   16 & 2\,792\,724 & 1\,706\,934 & 15\,225\,302 & 9\,317\,161\\
   18 & 21\,431\,970 &  13\,072\,764 & &   80\,130\,487 \\
\end{tabular}
\caption{Maximum number of states in the transfer matrix algorithm for an $L \times W$ rectangle with
$W \ge L$. Here $N_1$ (resp.\ $N_2$) refers to the time evolution before (resp.\ after) the addition of
the centre point. The corresponding primed quantities, $N'_1$ and $N'_2$, are for the improved algorithm
that exploits reflection symmetries.}
\label{tab:states}
\end{center}
\end{table}

It is quite simple to implement these changes. To keep the bottom boundary empty, it suffices to modify
the stage {\tt Start} so that initially the system is simply in the completely empty state (with weight $1$).
To keep the left boundary empty, we modify the {\tt Left} operation so that the leftmost half-edge cannot
acquire the {\tt SIDE} label. The maximum number of states encountered in the time evolution are compared
in Table~\ref{tab:states} for the original and the improved algorithms. It is seen that the improvement
diminishes the number of states by a factor that is asymptotically $\simeq 1.65$.

\section{Full hitting distribution}

\begin{figure}
\begin{center}

\begin{tikzpicture}[scale=0.5,>=stealth]

 \draw[help lines] (0,0) grid (4,6);
 \draw[black,thick] (0,0) -- (4,0) -- (4,6) -- (0,6) -- cycle;
 \draw[fill] (2,3) circle (1ex);
 \foreach \xpos in {1,2,3}
  \draw[blue,line width=2pt] (\xpos,0) -- (\xpos,6);
 \foreach \ypos in {1,2,3,4,5}
  \draw[blue,line width=2pt] (0,\ypos) -- (4,\ypos);
 \draw[fill] (2,3) circle (1ex);
 \draw (2,0) node[below]{(a)};

\begin{scope}[xshift=5.0cm]
 \draw[help lines] (0,0) grid (4,6);
 \draw[black,thick] (0,0) -- (4,0) -- (4,6) -- (0,6) -- cycle;
 \draw[fill] (2,3) circle (1ex);
 \foreach \xpos in {1,2,3}
  \draw[blue,line width=2pt] (\xpos,1) -- (\xpos,6);
 \foreach \ypos in {1,2,3,4,5}
  \draw[blue,line width=2pt] (1,\ypos) -- (4,\ypos);
 \draw[fill] (2,3) circle (1ex);
 \draw (2,0) node[below]{(b)};
\end{scope}

\begin{scope}[xshift=10.0cm]
 \draw[help lines] (0,0) grid (4,6);
 \draw[black,thick] (0,0) -- (4,0) -- (4,6) -- (0,6) -- cycle;
 \draw[fill] (2,3) circle (1ex);
 \foreach \xpos in {1,2,3}
  \draw[blue,line width=2pt] (\xpos,1) -- (\xpos,5);
 \foreach \xpos in {2,3}
  \draw[blue,line width=2pt] (\xpos,5) -- (\xpos,6);
 \foreach \ypos in {1,2,3,4,5}
  \draw[blue,line width=2pt] (1,\ypos) -- (3,\ypos);
 \foreach \ypos in {3,4,5}
  \draw[blue,line width=2pt] (3,\ypos) -- (4,\ypos);
 \draw[fill] (2,3) circle (1ex);
 \draw (2,0) node[below]{(c)};
 \draw (1,6) node[above]{-1};
 \draw (2,6) node[above]{0};
 \draw (3,6) node[above]{1};
 \draw (4,1) node[right]{-2};
 \draw (4,2) node[right]{-1}; 
 \draw (4,3) node[right]{\ 0};
 \draw (4,4) node[right]{\ 1};
 \draw (4,5) node[right]{\ 2}; 
\end{scope}

\end{tikzpicture}
 \caption{Boundary conditions that exploit the reflection symmetries of the lattice and improve the
 efficiency of the computations. The full lattice (a) can be reduced by depleting its bottom and left
 boundaries (b), and further depleting one half of the right and top boundaries (c). The coordinate
 $c_x$ (resp.\ $c_y$) labelling the horizontal (resp.\ vertical) exit point is defined in panel (c).}
  \label{fig:lattice}
\end{center}
\end{figure}
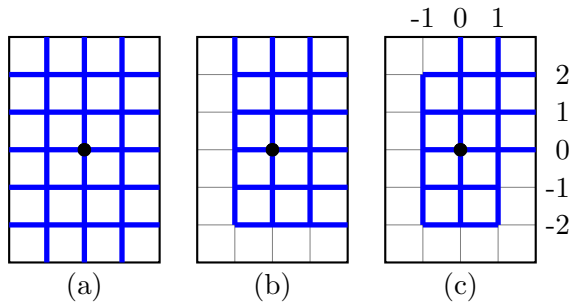

We have also developed a refined enumeration scheme in which the full hitting distribution is
recorded. The relevant geometry is shown in Figure~\ref{fig:lattice}. We wish to obtain separately
the generating functions for SAWs starting at the centre of the rectangle and exiting at any prescribed point
on its boundary. Due to the reflection symmetries it is enough to consider exits on the right and top
boundaries. Each exit point on the right (resp.\ top) boundary can be labelled by its vertical
(resp.\ horizontal) coordinate $c_y$ (resp.\ $c_x$) with respect to the centre of the rectangle, as shown in Fig.~\ref{fig:lattice}c.
Again invoking  the reflection symmetries, the generating functions only depend on $|c_x|$ and $|c_y|$,
so it is enough to consider the cases $c_x \ge 0$ and $c_y \ge 0$. Therefore the full lattice (Fig.~\ref{fig:lattice}a)
can effectively be replaced, first by a lattice with empty bottom and left boundaries (Fig.~\ref{fig:lattice}b), and
next by a lattice with half of the top and right boundaries empty as well (Fig.~\ref{fig:lattice}c).

The changes to the original program when going from the lattice shown in Fig.~\ref{fig:lattice}b to that in 
Fig.~\ref{fig:lattice}c are completely analogous to those described in the preceding section when going
from the lattice shown in Fig.~\ref{fig:lattice}a to that in Fig.~\ref{fig:lattice}b.

The only other change required to obtain the full hitting statistics is to register the coordinate ($c_x$ or $c_y$)
of the exit point. This is done by including one extra integer in the characterisation of each state. This integer,
initially set to zero, in unchanged in most of the evolution process, except at the unique stage where an occupied
monomer hits the boundary. At that stage it is set to the coordinate value of the exit point. To be precise,
this happens either when a {\tt SIDE} or {\tt BOTTOM} label is attached to the state, or when then {\tt CENTRE}
label hits the boundary.

The results for our usual example of the $L \times W = 4 \times 6$ rectangle are as follows:
\begin{eqnarray}
 c_x = 0 & : & x^3 + 6 x^5 + 6 x^7 + 10 x^9 + 22 x^{11} + 14 x^{13} \nonumber \\
 c_x = 1 & : & 3 x^4 + 6 x^6 + 8 x^8 + 15 x^{10} + 25 x^{12} + 14 x^{14} \nonumber \\
 c_y = 0 & : & x^2 + 2 x^4 + 4 x^6 + 10 x^8 + 8 x^{10} + 8 x^{12} + 6 x^{14} \label{eq:full-hitting} \\
 c_y = 1 & : & 2 x^3 + 3 x^5 + 7 x^7 + 10 x^9 + 14 x^{11} + 15 x^{13} \nonumber \\
 c_y = 2 & : & 3 x^4 + 6 x^6 + 8 x^8 + 15 x^{10} + 25 x^{12} + 14 x^{14} \nonumber
\end{eqnarray}
Summing over $c_y$, and multiplying by two to take into account both the left and the right
boundaries, we recover (\ref{eq:LR}). Similarly, summing over $c_x$ we recover (\ref{eq:BT}).
Another general feature brought out by (\ref{eq:full-hitting}) is that all powers of $x$ within a given
generating function must have the same parity. This parity can be inferred from the length of the
shortest SAW to the relevant boundary, i.e., it is the same parity as that of $\frac{W}{2}+c_x$
(resp.\ $\frac{L}{2}+c_y$).

Finally, we remark that a SAW attaining
the internal vertex $v_{\rm int}$ closest to the upper right corner of the rectangle must necessarily exit
at one of the two neighbouring vertices on the boundary, i.e., contribute either to the generating
function with $c_x = L/2-1$ or the one with $c_y = W/2-1$. Conversely, any SAW hitting one
of these two boundary points must necessarily pass through $v_{\rm int}$. The two generating
functions are thus equal, and this general result is of course brought out by (\ref{eq:full-hitting}).

\begin{table}
\begin{center}
\begin{tabular}{r|rr}
 $L$ & $N''_1$ & $N''_2$ \\ \hline
     4 & 8 & 49 \\
     6 & 42 & 665 \\
     8 & 254 & 7\,879 \\
   10 & 1\,670 & 86\,165 \\
   12 & 11\,596 & 898\,727 \\
   14 & 83\,670 & 9\,097\,463 \\
\end{tabular}
\caption{Maximum number of states in the full hitting statistics algorithm for an $L \times W$ rectangle with
$W = 2L$. Here $N''_1$ (resp.\ $N''_2$) refers to the time evolution before (resp.\ after) the addition of
the centre point. These numbers should be compared with those in Table~\ref{tab:states}.}
\label{tab:states-full}
\end{center}
\end{table}

The need to register the exit coordinate obviously increases the number of states used in the
computation. One might expect that roughly $W/2$ times more states would be needed. Fortunately,
the additional constraints imposed on the boundary half-edges ameliorates the situation somewhat.
The maximum number of states needed by the full hitting statistics algorithm are shown in Table~\ref{tab:states-full}.

\section{Analysis of data}

In Table \ref{table:aspect2} we give the ratio $R_n(2)$ of the number of walks hitting the long side to that of the number of walks hitting the short side for a rectangle with aspect ratio 2. 

\begin{table}[ht]
\caption{Ratio of number of walks first hitting long side to the number hitting the short side for a $n \times 2n$ rectangle, $R_n(2).$}
\centering
\begin{tabular}{c c  }

\hline
\hline 
$n$ & Ratio $R_n(2)$ \\ [0.5ex]
\hline

 2  &4.6381585303417408684303075667444130488805022010318359737078706077696\\
 4  &4.5626403997998714832892051885980313566654043362375413463572606810903\\
 6  &4.5737694425659079263885691980864259391038701844156223313634364418527\\
 8  &4.5816876116611105157124235041836861179217970948263182063453368731855\\
10 &4.586835597801674598789736657813550866717170851432063193366000200424\\
12  &4.590372957831729906013721282115051244422423169272971142266883828004\\
14  &4.592938409379941423415363287957199443993269739941033366762062737428\\
16  &4.594881296788588474819955070086767540896832163815763639405593859180\\
18 & 4.5964037339238675392443305096632781995673848727512239850106566816295\\ [1ex]
\hline
\end{tabular}
\label{table:aspect2}
\end{table}

This has been calculated by evaluating the ratio of the two ``generating functions" (which are of course polynomials, as we are dealing with finite lattices) at $x_c.$ For the square lattice $x_c$ is not exactly known, but the best numerical estimate, $x_c=0.37905227774965(13)$ \cite{C11}, is in agreement with the mnemonic $581x^4+7x^2-13=0$ \cite{C93}, and it is the appropriate root of this polynomial equation that we are using. Our results are insensitive to the uncertainty in the best numerical estimate of $x_c.$

This ratio is seen to be monotonically decreasing, and a plot of the data suggests behaviour of the form $R_n(r) \sim c + d/n^\theta.$ In order to estimate $\theta,$ we notice that a log-log plot of $n$ against $R_n(2)-R_{n-1}(2)$ has gradient $-2.0 \pm 0.08.$ This implies $\theta = 1 \pm 0.08,$ and in our subsequent analysis we assume $\theta=1.$ We then extrapolated the sequence $\{R_n(2)\}$ by the Bulirsch-Stoer algorithm \cite{SB80}, with the free parameter in that algorithm set to reflect $\theta = 1.$ The analysis is shown in Table \ref{table:extrap}.

\begin{table}[ht]
\caption{Extrapolation of the data in Table \ref{table:aspect2} by the Bulirsch-Stoer algorithm}
\centering
\begin{tabular}{c c c c c c c }
\hline
\hline 

 4.589060& 4.694672&4.609872 & 4.608214  & 4.608850 &     4.609984 &  4.609630       \\                                                                                   
 4.600917&4.615954&4.608989 & 4.608787 &  4.608949 &  4.609718 & \\
 4.604615&4.611393 &4.608899 & 4.609077 & 4.608726& &\\
 4.606178&4.610139 &4.608975 & 4.609228 & & &\\
 4.606991&4.609677 &4.609086 & & & &\\
 4.607480&4.609489 & & & & &\\
 4.607806& & & & & &\\ [1ex]

\hline
\end{tabular}
\label{table:extrap}
\end{table}

 In this way we estimate the limiting ratio as $4.6096 \pm 0.0002.$

We also analysed the data by several other appropriate extrapolation algorithms. The Brezinski $\theta$ algorithm \cite{G89} gives an increasing sequence of estimates for the large $n$ limit of the ratios, sugesting that the limit $\lim_{n \to \infty} R_n(2) > 4.6094.$ Levin's $u$-transform  \cite{G89} gives a monotone decreasing sequence of estimates, suggesting that $\lim_{n \to \infty} R_n(2) < 4.6097.$ Neville tables \cite{G89} also give a monotone increasing sequence of estimates, suggesting that $\lim_{n \to \infty} R_n(2) > 4.6090.$ These are all consistent with our estimate from the Bulirsch-Stoer algorithm.  This gives $b = 0.62501 \pm 0.00003,$ or $\kappa=2.66664 \pm 0.00007.$

In  Table \ref{table:aspect10} we give the corresponding, slightly shorter data set, for walks in a rectangle of aspect ratio 10. By a similar analysis, we estimate the limit to be $R_n(10)=14955 \pm 15.$ From (\ref{asympt}) this implies $b = 0.62496 \pm 0.000066,$ or $\kappa = 2.66675 \pm 0.00015.$

\begin{table}[ht]
\caption{Ratio of number of walks first hitting the long side to the number hitting the short side for an $n \times 10n$ rectangle, $R_n(10).$}
\centering
\begin{tabular}{c c  }

\hline
\hline 
$n$ & Ratio $R_n(10)$ \\ [0.5ex]
\hline

4 &  14006.18549331435655361766127203880086492399593549150863381964074817\\
6 &   14245.30058306730412888413593536062491411218703615652940541335814143\\
8  & 14391.29165062927743781254092745976155303961361728983232311816771246\\
10 &  14487.00740644606426682399787816797507371639547391783093463818418043\\
12  & 14554.35495421800301499956345001503940598563417509710466751030105068\\
14 & 14604.3566053623407864953033765640802317538963314516614175885263116878\\ [1ex]
\hline
\end{tabular}
\label{table:aspect10}
\end{table}

\begin{figure}[ht!]
\begin{center} 
\includegraphics[width=8cm]{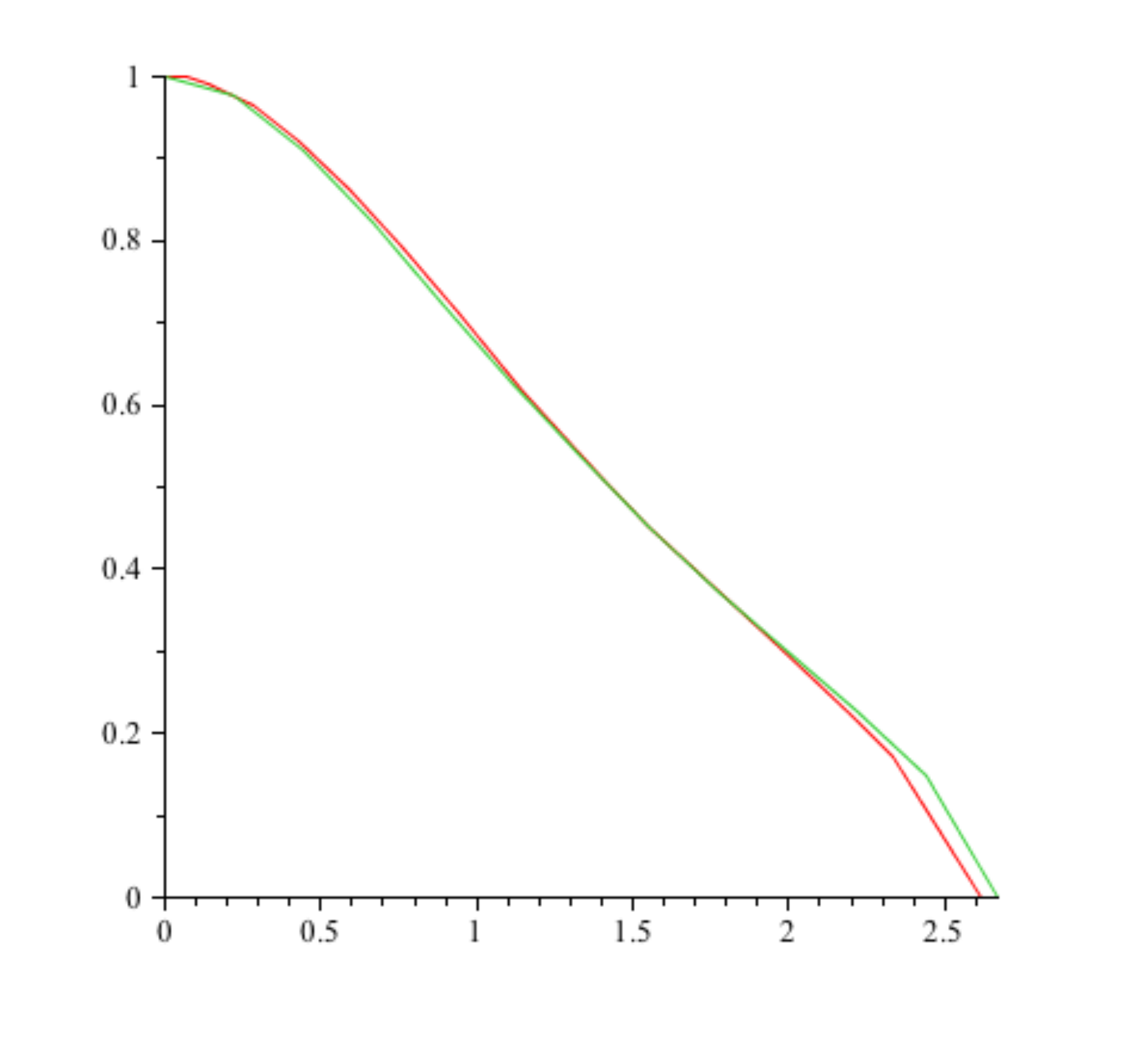}
\caption{Hitting density distribution function for a $12 \times 24$ rectangle (green) compared to the theoretical prediction in the scaling limit with $b=5/8$ (red).}\label{fig:hit-dens} 
\end{center}
\end{figure}

As well as calculating the ratio of the number of walks hitting the long side to that of the number of walks hitting the short side of a rectangle with aspect ratio 2, we have also calculated the distribution of hitting densities along the long side of such a rectangle. For the case of a $14 \times 28$ rectangle, the data is shown in  Table \ref{table:hitden}. From (\ref{cisaw}) it follows that the (unnormalized) probability density
satisfies
$$\rho_{f(D)}(f(z)) \propto \rho_D(z)|f'(z)|^{-b},$$ where $$f'(z)=(1-z^2)^{-1/2}(\alpha^2-z^2)^{-1/2}$$ follows from the Schwarz-Christoffel transformation previously discussed, and the unrenormalized hitting density along the real axis is $(z^2+\alpha)^{-b},$ as follows from (\ref{cirw})--(\ref{cisaw}), and is described in detail in \cite
{GK13}. So the hitting density along the long side of the rectangle is (up to a multiplicative constant),
$$(z^2+\alpha)^{-b}(1-z^2)^{b/2}(\alpha^2-z^2)^{b/2},$$ where the corresponding ordinates follow from the Schwarz-Christoffel transformation, so that the ordinates $(0,1/14, \ldots , 13/14)$ transform as
$$n/14 \to \int_0^{n/14} \frac{1}{\sqrt{1-t^2}}\frac{dt}{\sqrt{\alpha^2-t^2}}=\frac{1}{\alpha}{\bf F}\left ( \frac{n}{14},\frac{1}{\alpha} \right ),$$ where ${\bf F}$ is the incomplete elliptic integral of the first kind.
In figure \ref{fig:hit-dens} we show the predicted hitting density data for a rectangle of aspect ratio 2 together with the measured data for SAWs in a $14 \times 28$ rectangle. The enumeration data has been scaled linearly to make the origin and end-points coincide. Despite the fact that a $14 \times 28$ rectangle would seem to be rather far from the scaling limit, the agreement is remarkably good.

\begin{table}[ht]
\caption{Hitting distribution for a $14 \times 28$ rectangle. The third column gives the (unnormalized) fraction of walks hitting the long boundary at $\pm n/14,$ where $n$ is given in column 1 and $n/14$ in column 2.}
\centering
\begin{tabular}{c c c }

\hline
\hline 
$n$ & $n/14$ & Hitting number \\ [0.5ex]
\hline
0 &0.000000000000000000000000000000&  0.372179965985066951892060536379 \\
1 &0.071428571428571428571428571428 & 0.365592202518487166323769503648\\
2 &0.142857142857142857142857142857  &0.347584595139263204799699150831\\
3 &0.214285714285714285714285714286  &0.321757459504506388412145573956\\
4 &0.285714285714285714285714285714  &0.291774774902074605895271622535\\
5 &0.357142857142857142857142857143  &0.260478252386315460357350977416\\
6 &0.428571428571428571428571428571  &0.229709762715325246279406382177\\
7 &0.500000000000000000000000000000  &0.200461104152704918984187699069\\
8 &0.571428571428571428571428571429  &0.173099730010024355136783990185\\
9 &0.642857142857142857142857142857  &0.147556785152232511539056308909\\
10 &0.714285714285714285714285714286  &0.123439559572782004646612438472\\
11 &0.785714285714285714285714285714  &0.100052601022725874927123950740\\
12 &0.857142857142857142857142857143  &0.076292442893068500584108376384\\
13 &0.928571428571428571428571428571  &0.050321628007048540057861695734\\ [1ex]

\hline
\end{tabular}
\label{table:hitden}
\end{table}



\section{Conclusion}
We first calculated the ratio of the number of SAWs starting at the centre of a rectangle and hitting the end to that of the number of SAWs starting at the centre of a rectangle and hitting the sides, for rectangles of a given aspect ratio. Then by comparing this to the calculated value \cite{GK13} assuming that the scaling limit of SAWs is given by $SLE_\kappa,$ we estimated $\kappa$ for rectangles of aspect ratio 2 and 10. In the former case we found  $\kappa=2.66664 \pm 0.00007,$ and in the latter case we found $\kappa=2.66675 \pm 0.00015.$ This would appear to be the strongest numerical support yet for the widely held belief that the scaling limit of SAWs is given by $SLE_{8/3}.$

Furthermore, we have also calculated the distribution of hitting densities along the long side of such a rectangle and compared this with the actual distribution for a $14 \times 28$ rectangle. Visually, the agreement between the plots of the two distributions is seen to be quite persuasive.

\newpage

\section*{Acknowledgements}  This work was supported by the Australian Research Council through grant DP120100939  (AJG and JLJ), and by the Institut Universitaire de France and Agence Nationale de la Recherche through grant ANR-10-BLAN-0414 (JLJ). AJG would like to thank Larry Glasser for a lesson in asymptotic expansions, Tom Kennedy for a lesson in conformal mapping, and both authors would like to gratefully acknowledge the support of the Simons Center for Geometry and Physics at SUNY Stony Brook where this work was, in part, carried out.

\end{document}